# Characteristic nanoscale deformations on
# large area coherent graphite moiré

N. Sarkar[1*], P.R. Bandaru,[1,3] R.C. Dynes[2,3*]

[1]Department of Mechanical Engineering, [2]Department of Physics,

[3]Program in Materials Science,

University of California San Diego, La Jolla, California 92093-0411, USA

Highly oriented pyrolytic graphite (HoPG) may be the only known monatomic crystal with the ability to host naturally formed moiré patterns on its cleaved surfaces, which are coherent over micrometer scales and with discrete sets of twist angles of fixed periodicity. Such an aspect is in marked contrast to twisted bilayer graphene (TBG) and other multilayered systems, where the long range coherence of the moiré is not easily maintained due to twist angle disorder. We investigate the electronic and mechanical response of coherent graphite moiré patterns through inducing external strain from STM tip-induced deformation. Consequently, unique anisotropic mechanical characteristics are revealed. For example, a lateral widening of one-dimensional (1D) domain walls (DWs) bridging Bernal (ABA) and rhombohedral (ABC) stacking domains (A, B and C refer to the atomic layer positioning), was indicated. Further, *in situ* tunneling spectroscopy as a function of the deformation indicated a tendency towards increased electrical conductance, which may be associated with a higher density of electronic states, and the consequent *flattening* of the electronic energy band dispersion. Such features were probed across the DWs, with implications for strain-induced electronic modulation of the moiré characteristics.

The natural occurrence of periodic and coherent moiré patterns unaffected by defects on exfoliated HoPG surfaces over large areas may be associated with the energetically favorable nature of superlattice formation (1). Consequently, HoPG may be considered a reliable experimental platform for investigating the effects of intrinsic and extrinsic strain (2) and the related modulation of electronic energy dispersion(3). Such "soft" (4) graphite surfaces could be probed by STM topography and spectroscopy and reveal electromechanical characteristics beyond those observed on twisted bilayer graphene (TBG) and other 2D-like materials systems. While TBG moiré modulated through hetero-strain(5,6), electrical gating(7), tip-induced deformation(8) or hydrostatic pressure(9) are interesting for the possibilities of superconductivity(10) and electronic flat bands(11), it is well known(12) that such systems are yet prone to surface defect induced-incoherent moiré. Alternately, the moiré on atomically flat HoPG would be less influenced by such surface defects and can also be an ideal substrate for fabricating large scale coherent moiré patterns. The wider possibility of domain arrangements in graphite moiré, *i.e.,* AA, ABA, ABC, also yields a richer variety of topological configurations that may be investigated.

Atomic sheets constituting the HoPG are coupled by weak van der Waals (vdW) forces that allow cleavage of the constituent layers along the basal planes. Often, the partially lifted layers in cleavage reattach to the HoPG surface, with rotation, to form a moiré pattern as shown in **Fig. 1 (a)**. We have observed the occurrence of multiple discrete sets of moiré patterns of varying twist angles ($\theta_{twist}$) with corresponding periodicities. In contrast, in TBG synthesized through chemical vapor deposition(13) or dry transfer(12), incoherent periodicity in the moiré is nominally observed *i.e.,* a twist angle disorder, presumably due to the inevitable occurrence of defects(12) related to substrate roughness, polymer residues, wrinkles, *etc*. However, moiré on HoPG is less susceptible to such disorder, and defects such as wrinkles or steps may instead act as pinning centers yielding discrete changes in $\theta_{twist}$: **Fig. 1(a)** - *also see Figure S1 in Supplementary section S1*). The tilt boundaries marking the edges of the moiré superlattices are denoted by arrays of orange dots in **Fig. 1(a)** and seem displaced out of the plane indicating highly strained regions or *tilt boundaries* (*see Supplementary section S1*).

An instance of a graphite moiré pattern with $\theta_{twist}$ = 4.2° is shown in **Figs. 1(b)** and **1(c)**. The periodic occurrence of the AA domains (hill-like features), ABA domains (valley-like features) and the bridging DWs is illustrated. The AA atom stacking is an unstable configuration due to direct placement of atoms on the individual layers on top of each other while the ABA/Bernal stacking is more stable with an atom from the top layer occupying the spaces between the atoms of the lower layer or *vice versa*. In this work, STM tip-induced surface deformation(4,14) of moiré patterns was used for revealing local vdW bonding strength variability(1,15) across the moiré landscape as well as elastic mechanical deformation characteristics perpendicular to the surface, as indicated through the schematic of **Fig. 2(a)**. The STM tip was scanned in the direction: AA → DW → AA: *blue dashed line,* or along AA→ABA→ AA: *black dashed line*, as indicated in **Fig. 2(b)**. The related topography is shown in **Fig. 2(c)** and **Fig. 2(d)** respectively as a function of the tunneling current ($I_{tip}$).

The observed high (/low) surface deformation amplitude in the AA (/ABA) domains could be ascribed to the associated weaker (/stronger) vdW bonding while the bonding in the DW region itself appears intermediate to that relative to the AA (/ABA) stacking. The deformation amplitudes of the domains measured from **Figs. 2(c)** and **2(d)** are indicated in **Fig. 2(e)** - as a function of the $I_{tip}$ and $R_{gap}$ (tunneling gap resistance = $V_{sample}/I_{tip}$, with $V_{sample}$ as the applied bias voltage on the HoPG). At a reduced tip-sample gap, *i.e.,* with lower $R_{gap}$, there is considerable outward deformation/*bulging* out of the unstable AA stacking by ~ 1.1 nm whereas the ABA regions are much less deformed, *i.e.,* at ~ 0.3 nm, due to stable stacking configurations and stronger bonding. The high amplitude of these deformations dominate over the electronic states variation in topography maps. Hence, a topographic map of the graphite moiré could be considered a map of its interlayer vdW strength. Tip-induced deformations on "soft" HoPG(4) would be of much higher magnitude relative to TBG(8,14) due to the larger influence from the substrate in the latter case. To guard the tip shape and geometry during deformation, small tunneling resistances (~ kΩ) corresponding to small gap distances were avoided. In addition to the surface deformations of domains, tip-induced atomic forces also pull individual atoms. The *inset* in **Fig. 2(d)** indicates enhanced atomic corrugations on AA domains compared to those on ABA sites and are plotted in **Fig. 2(f)**. It was observed here that the AA domains appear less rigid relative to the ABA domains.

Any external strain induced by the STM tip would be expected to deform the moiré surface features anisotropically, with deformation perpendicular to the surface, expected to be larger than in the surface plane(1,8,11). We show such asymmetry in the mechanical characteristics through monitoring STM tip-induced deformation as a function of $\theta_{twist}$. The moiré patterns with $\theta_{twist}$ of 4.2° and 0.5° are shown before deformation in **Figs. 3(a)** and **3(b)** and after deformation (through an increased $I_{tip}$) in **Fig. 3(c)** and **Fig. 3(d),** respectively. The transition of the original moiré pattern to a deformed one with tip-induced deformation was reversible over the chosen range of operation. For $\theta_{twist}$ = 4.2°, the AA regions ("hills") relatively occupy the same surface area in topography mapping as the ABA regions ("valleys") before and after the deformation: **Fig. 3(a)** and **Fig. 3(b)**. The relative surface area occupied by the moiré domains is a measure (16) of the local strain energy balance. However, with $\theta_{twist}$ reduction from 4.2° to 0.5°, it was observed that (i) the periodicity of the AA domains has increased by an order of magnitude from ~ 3 nm to ~ 30 nm, (ii) triangular patterns of ABC and ABA domains are dominant, and there is a (iii) pronounced stretching of DWs into linear structures: **Fig. 3(a)** and **Fig. 3(c)**. Domain Walls are less stable transition regions with a gradient in stacking configuration(13) occupying narrow spaces between relatively stable ABA and ABC stacked domains. Consequent to deformation in **Fig. 3(d)**, the 1D DWs stretch laterally into a 2D area shrinking the triangular ABC and ABA domains. Such longitudinal and lateral stretching behavior of DWs may be related to their soliton-like nature(13). Indeed, DWs have previously been shown (17) to be elastic string-like and may be correspondingly deformed with the STM tip depending on the direction of approach towards the ABC(/ABA) boundary. Such behavior is obscured(8) in TBG systems due to the lack of rhombohedral ABC domains.

To further probe the deformation characteristics along the direction perpendicular to the surface, the topography in the undeformed and deformed moiré patterns is compared in **Fig. 3(e)**. It was observed that prior to deformation (*black trace*), the ABC domains have a higher plateau compared to the ABA region. However, subsequent to deformation (*blue trace*), the AA domains are most affected out of plane, more than the ABC and the ABA domains. The ABA (*red label*) and ABC (*yellow label*) domains shrink in surface area while the DW (*green label*) region enlarges. The average deformation amplitude across the moiré surface, as a function of $\theta_{twist}$, is indicated in **Fig. 3(f)** where the error bars correspond to atomic corrugations (*See Supplementary S2*). A lower out-of-plane surface deformation is observed for smaller $\theta_{twist}$. However, along the surface, an opposite trend was observed, as may be seen from **Fig. 3(c) - 3(d).** Such anisotropic response, *i.e.,* of increased flexibility perpendicular to the surface and diminished flexibility coplanar with surface is observed when $\theta_{twist}$ is < 1° (*See Supplementary S3*). The amplification of the moiré periodicity at the lower $\theta_{twist}$ allows an enlarged metastable ABC domain(1) which imparts an increased flexibility to the moiré domains on the surface.

Such domain reconfiguration - under external strain – was further investigated. The gradual evolution of the lateral stretching of DWs with increased mechanical deformation, *i.e.,* increased $I_{tip}$ increment (at a fixed $V_{sample}$) is shown in **Fig. 4(a)→(c)**. The independence on $V_{sample}$ (at fixed $I_{tip}$) implies minimal electronic influences(18) *(See Supplementary Section S4)*. This deformation mechanism is consistent for several observed $\theta_{twist}$ < 1° *(See Supplementary Section S3)* indicating purely mechanical deformation effects. Further, the deformation pattern was independent of the tip-scanning direction. The schematic of the undeformed and deformed moiré is mapped in **Fig. 4(d)**. The three-fold rotational symmetry(1) of the alternating ABC and ABA sites is maintained. The spatially pinned AA domain determines the intersections of all the DWs(13), which grow wider subsequent to deformation as observed in **Fig. 4(e)**. Along the DWs, there is a gradient in topography, as indicated by the arrows - from bright to dark- in **Fig. 4(e)** and may be ascribed to constituent electric polarizability(18). Indeed, the flexoelectric character(19) of DWs induced through strain gradients from STM tip-induced interactions could be indicated. These strain gradients are arranged similar(20) to the triangular networks of highly conductive 1D electronic channels(21) around AB/BA domains. Such a network have previously (22) exhibited Aharonov–Bohm oscillations whose periods correspond to one magnetic flux quantum threading through the enclosed triangular domain areas only in TBG morphologies, and such effects may also be investigated in deformed graphite moiré.

The electronic DW channels merge at every metallic AA node where they can get coupled to further demonstrate weak Shubnikov-de Haas oscillations(21). Such AA sites have also been previously(18) attributed to a gapless *flat-band* energy dispersion associated with a high density of electronic states (DOS). Further enhancement of the DOS with tip-induced local strain induced is shown in **Fig. 5 (a)** through an increased electrical conductance (*dI/dV*) probed on AA sites of $\theta_{twist}$ = 0.8°. The in-situ spectroscopy reveals the narrowing of the conductance peaks, with decreasing $R_{gap}$: **Fig. 5 (c)** bottom to top. Such spectroscopy modulation is reproducible on repeated tip pulling in/out of the AA stacked layers. However, there is no observed difference in I(V) spectroscopy on AB-stacked HoPG surfaces before, during and after tip-induced deformation.

The conductance modulation through atomic tip-induced strain[11] can also be applied on distinctive electronic peaks of other domains and DWs. Such characteristic bias-specific peaks have previously[23] been shown on ABA/ABC domains but not on graphite DWs. In TBG, the DWs have been associated[21,23] with edge states which were predicted[24] to be difficult to probe in graphite moiré. The long-range coherence of graphite moiré favors conductance probing of electronic states gradually across the domains, *e.g.,* a typical path (*blue arrow*), AA → ABC → DW → ABA, is shown in **Fig. 5(a),** on a graphite moiré pattern of $\theta_{twist}$ ~ 1.1°. Here, unique bias-specific peaks were observed for the domains and domains walls: **Fig. 5(b)**. For instance, the peak at ~100 mV was associated with the AA domain from the high conductance and expected large DOS[18]. The single peak subsequently evolves into a pair when the ABC domains are probed - with a peak separation/gap ($\Delta_{ABC}$) of ~ 50 mV. Such gap formation has been previously reported[23,25] along with its calculated band structure[23] shown in the top-left *inset* of **Fig. 5(b)**. The $\Delta_{ABC}$ gradually disappears as the ABA domain is encountered. Here, a parabolic band energy dispersion was predicted[23] with the calculated band structure - shown in the top-right *inset* of **Fig. 5(b)**. While $\Delta_{ABC}$ could be increased through electrical gating[26,27] in TBG, this would not be possible on graphite moiré[24]. Such an increased gap would be necessary for the existence of electrical field induced topological chiral edge modes at AB-BA domain walls[21,23]. However, in graphite moiré, the edge modes related to the ABA-ABC DWs would be relatively dispersive[24] and difficult to probe. The set of peaks at ~300 mV (*green arrows*) in **Fig. 5(b)** observed close to the boundary between the ABC and ABA domains could be associated with the corresponding DW. The peak at ~ 400 mV - seen on all the domains except ABA, suggests the possibility of a delocalized energy band[28].

The domain area modulations through atomic tip-induced strain would modify their electronic characteristics and suggest possibilities in the use of graphite moiré for highly sensitive and specific stress or pressure[9] sensors tunable at atomic scale. At a $\theta_{twist}$ = 1.1°, the graphite Moiré is expected and was indeed observed to have different electronic spectroscopy signatures from that in TBG arising from the possibility of ABA/ABC domains in graphite moiré, in contrast to the AB/BA configuration in TBG. Also see *Supplementary Section S5,* for the possibility of nematic phase formation (16,18) at such $\theta_{twist}$. The deformation characteristics of domains and DWs, as elucidated in our work, is shown to be related to the weak interlayer interactions for multiple layered HoPG. However, the stronger physico-chemical influence of the substrates on 2D stacks like TBG possibly would make it difficult to observe such deformation phenomena.

In summary, STM based structural and spectroscopic investigations on graphite moiré yields new observations and insights, not evidenced in bilayer graphene moiré such as: (i) naturally occurring coherent graphite moiré over micron scale with discrete/discontinuous change in $\theta_{twist}$, (ii) Crystallographic directions or $\theta_{twist}$ dependent mechanical deformation characteristics, (iii) lateral stretching of 1-D DWs into 2-D domains, mimicking soliton-like behavior, with strain gradients that may be related to electrical polarizability and indicating a topological transport network that may exhibit Aharonov-Bohm oscillations (iv) controlled relative surface area modulations of the moiré domains and DWs. The possibility of robust strain-induced modulation of electrical conductivity motivates the use of graphite moiré for sub-nanoscale displacement sensing devices.

## Methods:

## Materials

Highly oriented pyrolytic graphite (HoPG, ZYH grade from Advanced Ceramics, Inc) was freshly cleaved before loading into the STM chamber. The choice of low grade ZYH with micron sized grains favored the occurrence of surface defects like step heights or grain boundaries which would act as pinning sites for moiré pattern formation during cleavage.

## Electrical Measurements

A custom-built scanning tunneling microscope (STM) with a RHK controller at room temperature and atmospheric pressure was used. The imaging was done at a nominal sample bias of 0.1V and tunneling current of 10nA. The topography measurements were performed in standard constant current mode at ~0.5 Hz scanning frequency using a mechanically snipped Pt/Ir tip. For surface deformation experiments, the tip was brought closer to the sample by varying the tunneling currents from 10 nA to 100 nA. Tunneling spectroscopy measurements utilized a lock-in modulation of 3mV at 5kHz to the sample bias which was swept from -0.4V to 0.5V.

## Calibration and Data analysis

The reported topographic length scales in-plane (x, y) and out-of-plane (z) were calibrated through atomic lattice constants of graphite through atomic imaging of the surface and its monatomic step. The topography curves in Fig. **2(c-d)** have no z-offset and this offset was observed to increase with the increase in tunneling current. The surface deformation amplitudes of the topographic hills and valleys were averaged over five moiré unit cells and the atomic corrugation estimation were averaged over ten atomic unit cells. The error bars estimation for surface deformation amplitude measurement in Fig. **3(f)** is described in Supplementary Section S2. All the imaging and deformation measurements were done using the same tip. The spectroscopy curves in Fig. **5(c)** have been offset for clarity. All domains and their conductance curves are color coded: AA (pink), ABA (red), ABC (yellow) and DW (green).

Acknowledgements:

This work was supported by AFOSR Grant (FA9550-15-1-0218) and Army Research Office (AROW911NF-21-1-0041). The authors wish to thank Michael Rezin for the technical assistance; Prof. Shane Cybart, Uday Sravan Goteti and Hidenori Yamada for useful discussions.

Author contributions:

N. Sarkar did the experimental work and along with P.R. Bandaru and R.C. Dynes, wrote the paper. All analysis and discussion were under the supervision of R.C. Dynes and P.R. Bandaru.

Data availability:

The experimental data and its analysis in the paper and/or in the supplementary information is sufficient to support our conclusions. Additional data can be made available on request.

Competing interests:

The authors declare no competing interests.

# **Figures**

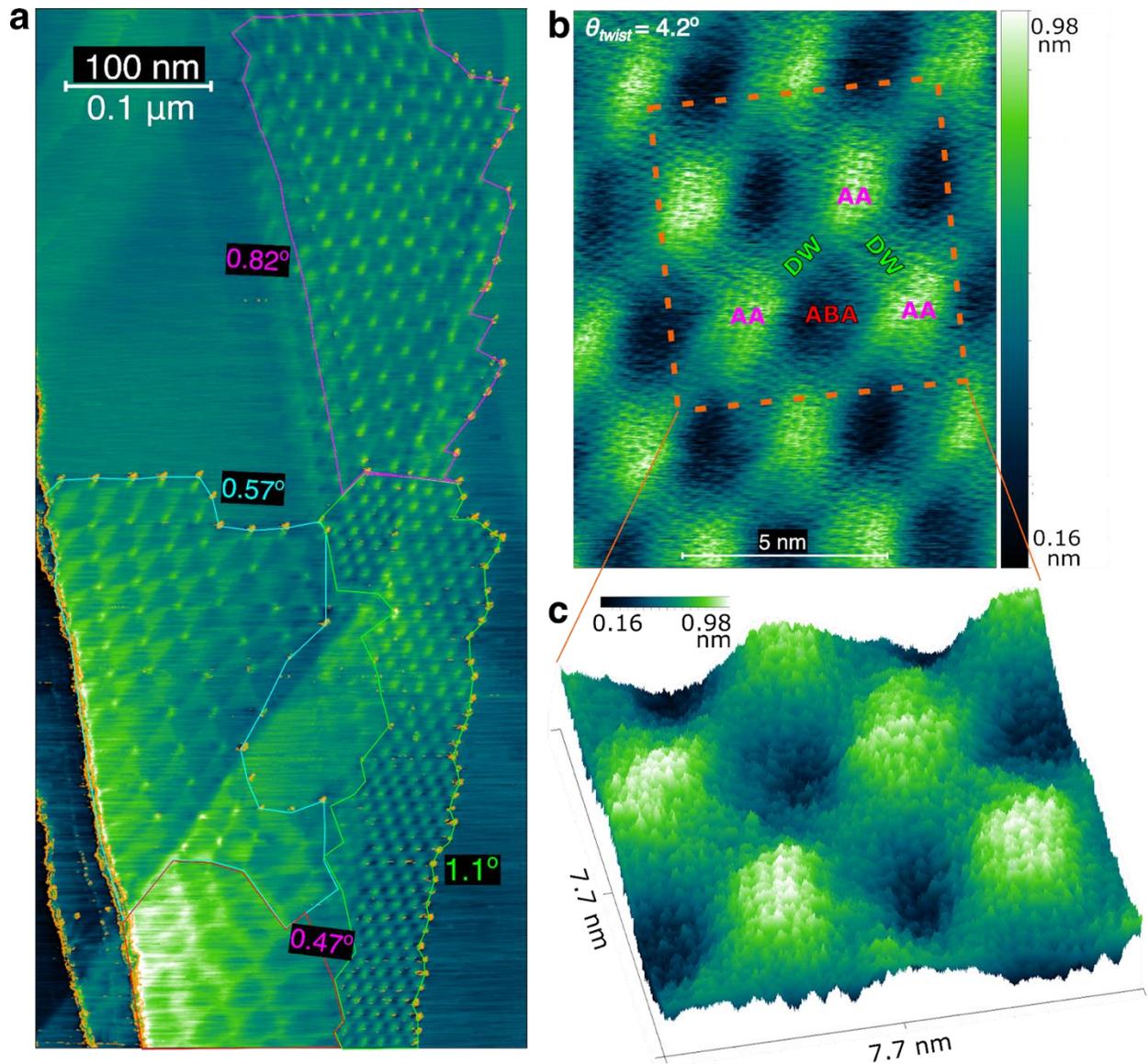

Figure 1: **Moiré patterns on graphite surface and related domain features. (a)** The occurrence of coherent moiré over micron scales with multiple twist angles ($\theta_{twist}$) in degrees. The orange dots at the edges are tilt boundaries. **(b)** The observed domain pattern, at a $\theta_{twist}$=4.2°, with the AA domains (*bright* hill-like features), ABA domains (*dark* valley-like features) and the bridging domain walls (DWs). **(c)** Atomically resolved three-dimensional topography of a region from **(b)**. The small dots correspond to individual atoms.

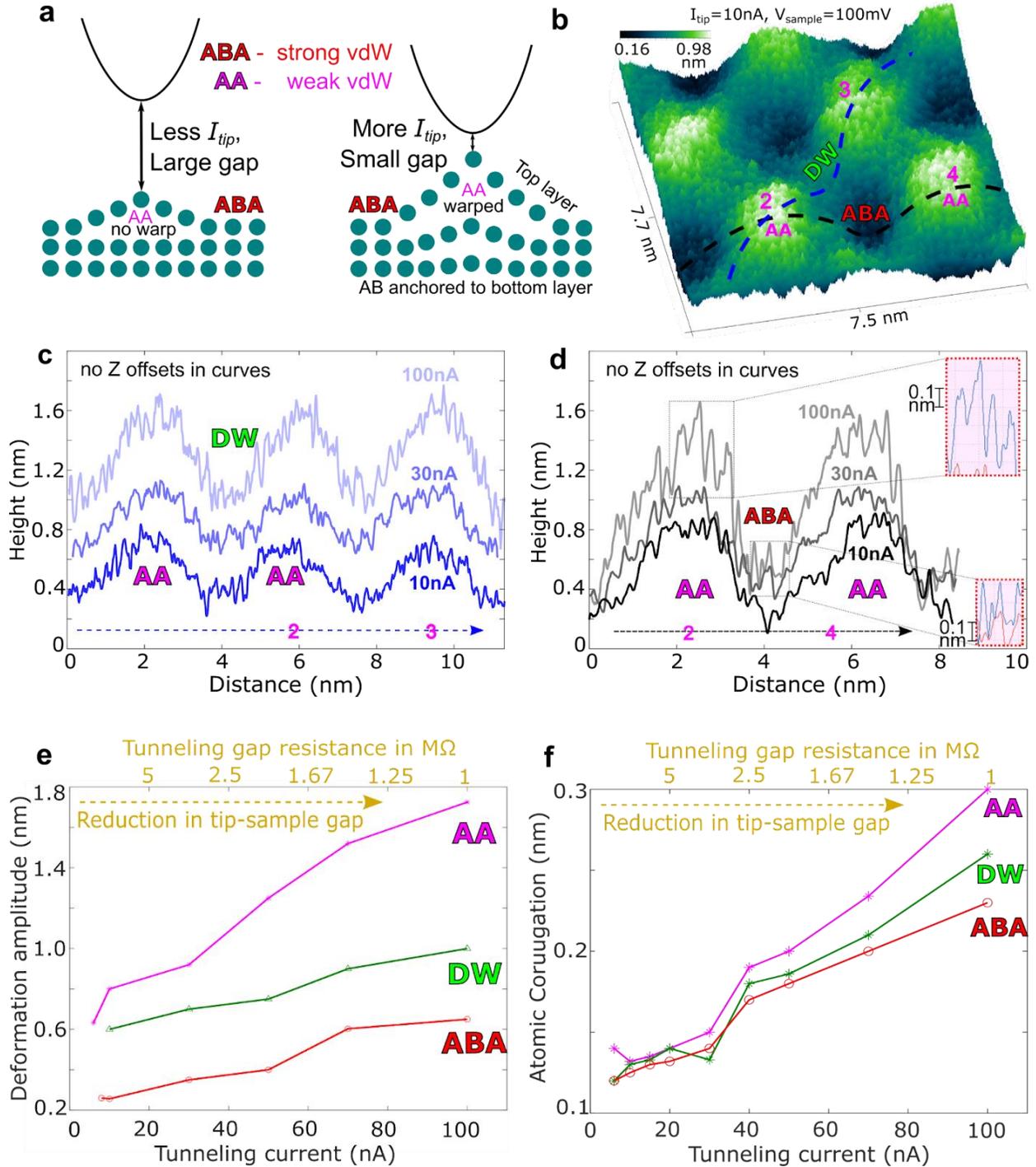

Figure 2: **Scanning Tunneling Microscopy (STM) based investigations of the deformation attributes, and van der Waals (vdW) bonding variability, in moiré pattern domains.**

**(a)** The contours of the Moiré domains being deformed by STM tip through large(/small) atomic forces monitored through large(/small) tunneling currents at smaller(/larger) tip-sample distance. **(b)** An atomically resolved topography of the moiré pattern, with $\theta_{twist}$ = 4.2°. Two distinct scan directions (AA→DW→AA, along the DW: *blue dotted line*, and AA → ABA → AA: *black dashed line*) are shown, with the corresponding atomic amplitude variations as a function of the $I_{tip}$ in (c) and (d), respectively. The *insets* in **(d)** indicate the atomic corrugations on a deformed sheet. A larger (/smaller) corrugation was observed in the AA(/ABA) domains, due to the weaker (/stronger) vdW bonding. The scans in **(c)** and **(d)** are not z-offset and consequently indicate surfaces being pulled out of plane. **(e)** A larger (/smaller) extent of deformation was seen for the AA (/ABA) domains with an intermediate extent for the DWs. **(f)** The magnitude of the individual atomic corrugation is higher on AA domains relative to the DW and ABA domains, due to the underlying bonding configuration.

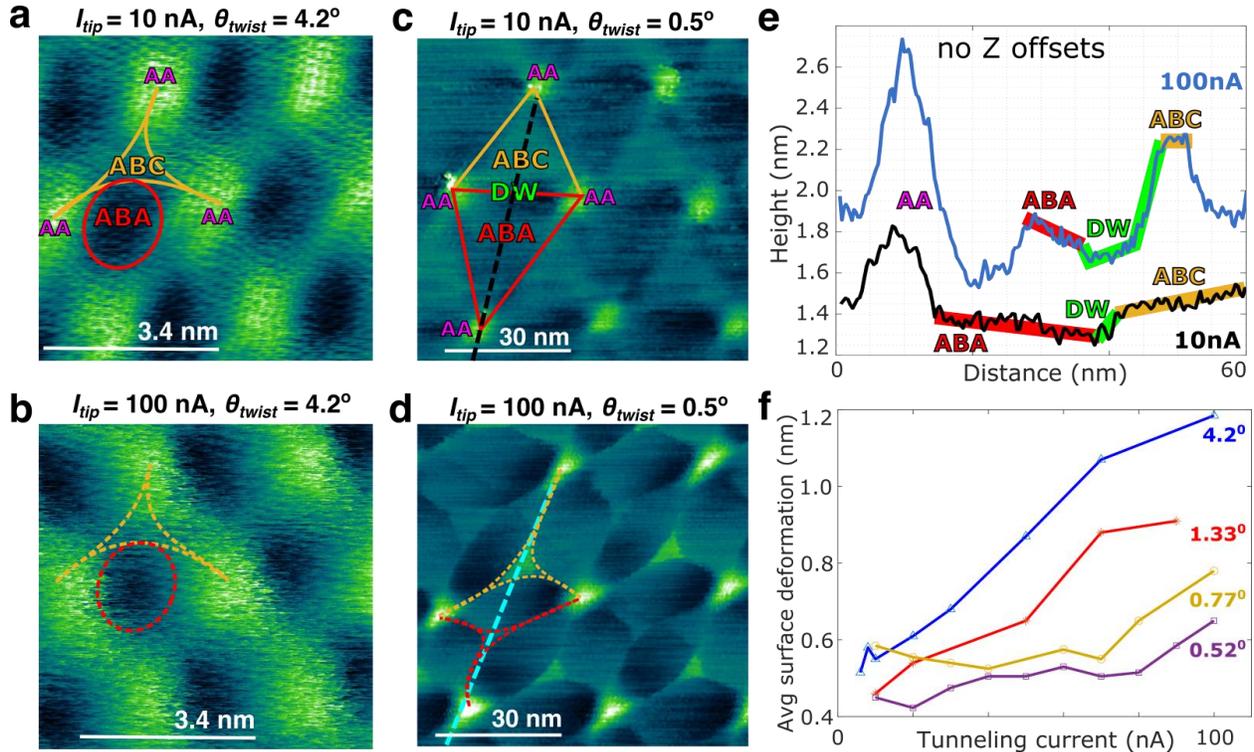

Figure 3: **Moiré twist angle dependent deformation.** The depiction of the moiré and constituent domains with $\theta_{twist}$ = 4.2° **(a)** before, and **(b)** after deformation – deduced from an increased $I_{tip}$ from 10 nA to 100 nA. Here, the AA domain periodicity is ~ 3.4 nm, and does not change appreciably. The ABA and ABC domains as well as the DW (*stretched triangle*) configuration is unaltered. **(c)** Alternately, with $\theta_{twist}$ = 0.5° the AA domain periodicity is significantly larger at ~ 30 nm. The triangular ABA and ABC domains are larger when compared to their size at a larger $\theta_{twist}$. The related DWs are straight and are **(d)** reconfigured subsequent to deformation, through an increased $I_{tip}$ from 10 nA to 100 nA, with an accompanying reduction in the size of the ABA and ABC domains. **(e)** The topography traced along the *black* and *blue* dotted lines in **(c)** and **(d)** as a function of $I_{tip}$ indicates an out of plane displacement for the domains in the order AA > ABC > ABA. **(f)** A comparison of the deformation of the moiré surface as a function of $I_{tip}$ at a given $\theta_{twist}$, indicates a greater magnitude at a lower $\theta_{twist}$. All images were measured with $V_{sample}$ is 100 mV. All domains are color coded: AA (*pink*), ABA (*red*), ABC (*yellow*) and DW (*green*).

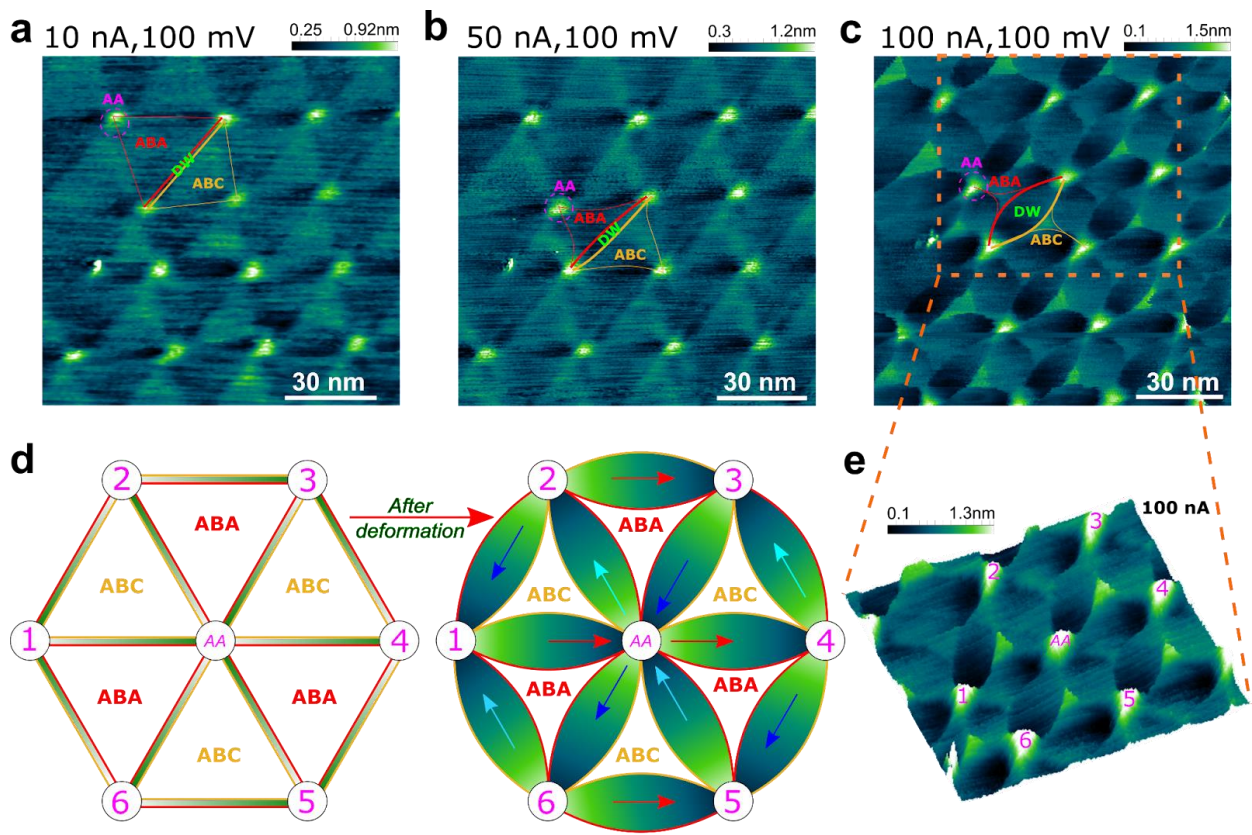

Figure 4: **The reconfiguration of the moiré domains, at $\theta_{twist} = 0.5°$, as a function of tunneling current and tip-sample bias. (a)** A reference domain and DW configuration at an $I_{tip} = 10$ nA, and $V_{sample} = 100$ mV. Keeping the $V_{sample}$ fixed, there is an increased DW stretching as the $I_{tip}$ is increased to **(b)** 50 nA, and **(c)** 100 nA. Bias independence of this pattern is shown in Supplementary S4 **(d)** The domain pattern exhibits a three-fold symmetry, with AA as an origin (*left*), and this symmetry is unaltered by tip-induced deformation (*right*). DWs grow wider and the gradient in topography along the DWs - indicated by the arrows from bright to dark may be ascribed to constituent electric polarizability and flexoelectric character. The clockwise/counter-clockwise network of DW electronic channels around ABA/ABC domains can exhibit Aharonov-Bohm effects like in TBG **(e)** A 3D topography map corresponding to deformed moiré pattern in **(c)**.

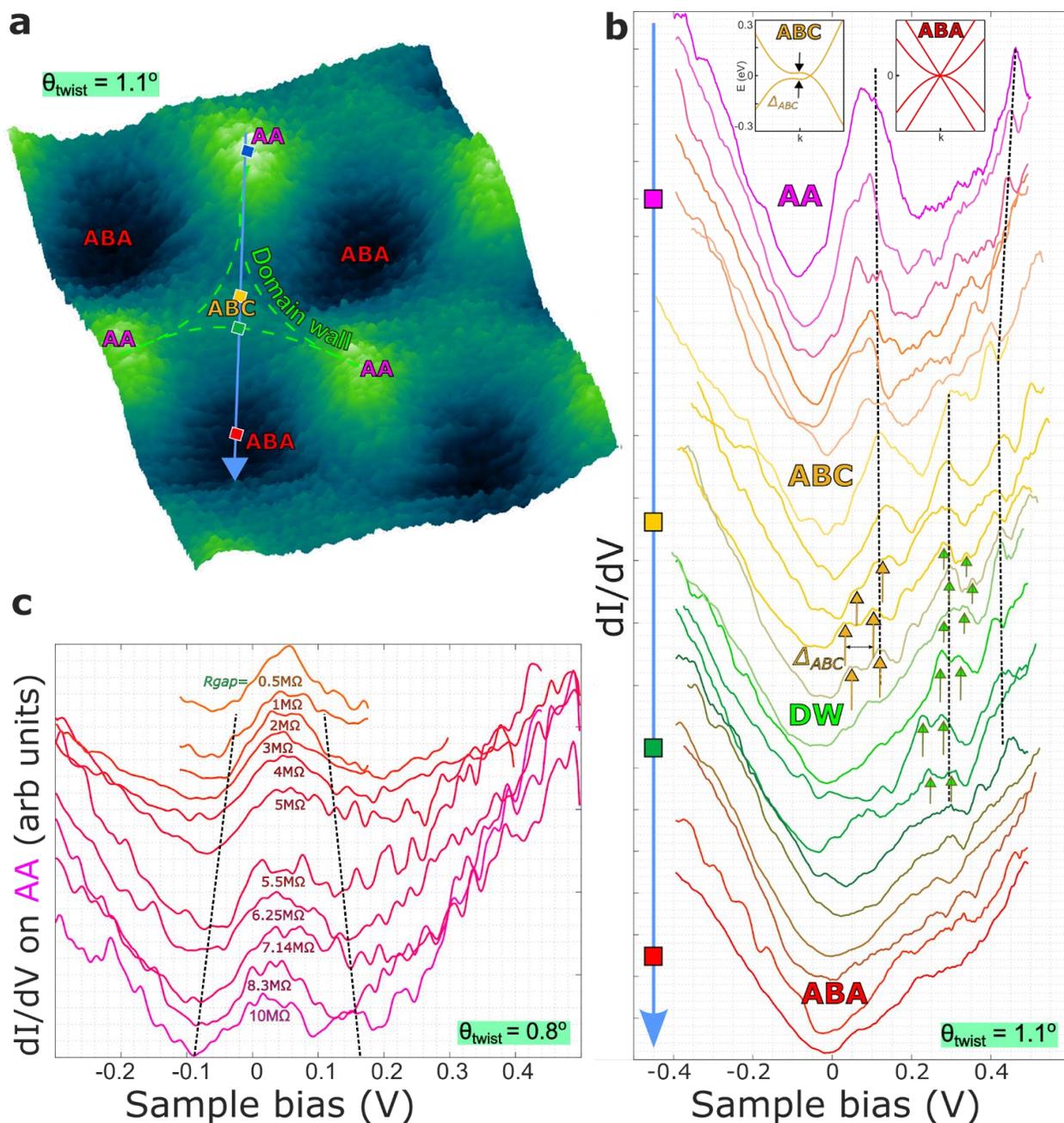

Figure 5: **Electrical conductance mapping across moiré domains.** **(a)** Moving with the STM tip along the *blue* arrow, on a moiré with $\theta_{twist}$ = 1.1° ($I_{tip}$=10nA, $V_{sample}$=100mV) across the domains: AA → ABC → DW → ABA, yields conductance traces as shown in **(b).** Here, the peak at ~100 mV was associated with the AA domain from the high conductance and expected large density of states, which evolves into a pair when the ABC domains are probed - with a peak separation/gap ($\Delta_{ABC}$) of ~ 50 mV. The related band structure is shown in the top-left *inset*. The $\Delta_{ABC}$ gradually diminishes as the ABA domain is encountered, with a parabolic band energy dispersion - shown in the top-right *inset*. The set of peaks at ~300 mV (*green arrows*) observed close to the boundary between the ABC and ABA domains could be associated with the corresponding DW. **(c)** For a moiré with $\theta_{twist}$ = 0.8°, the conductance on the AA domains increases with a reduced $R_{gap}$ (/increased $I_{tip}$) along with a decreased peak width. Such spectroscopy modulation is reversible on repeated tip pulling in/out of the AA stacked layers.

# Supplementary Information

N. Sarkar[1], P.R. Bandaru,[1,3] R.C. Dynes[2,3]

[1]Department of Mechanical Engineering, [2]Department of Physics,

[3]Program in Materials Science,

University of California San Diego, La Jolla, California 92093-0411, USA

## S1. Role of surface defects in forming moiré patterns

Formed moiré patterns on the graphite surface are usually observed after dry tape exfoliation. They have often been observed next to step heights where the graphite sheets have an open edge which would most likely be the edges where the sheets would be capable of being lifted by tape. Partially lifted sheets during the exfoliation occasionally may not get carried away by the tape and would settle back down with a small twist w.r.t the surface thereby forming a moiré pattern. Such patterns are found to be coherently outspread over microns or sub-microns. While the sheet settles down to form a coherent moiré, it can get pinned on defect regions where the twist angle can get slightly changed. This is called 'twist angle disorder' also shown in Fig. 1a where a 1D defect like wrinkle causes a small shift in twist angles. The variation in twist angles can be calculated from the change in moiré wavelength. Twist angle disorder is a common issue when dry transfer techniques are used due to the occurence of surface defects like substrate induced roughness, wrinkles, polymer residues, etc. The step edge on the left is possibly the edge where the sheet was partially lifted.

These patterns have the strongest strain density at the edges (termed 'tilt boundary') associated with endurance of the whole strained twisted sheet. The tilt boundaries are shown with colored lines in Fig. S1 (a) and can be observed to be topographically higher than the rest of the moiré sheet as shown in the 3-dimensional view: Ref. Fig. S1 (c). When multiple twist angles coexist next to each other with shared tilt boundaries as seen in Fig. S1 (a), a closer atomic look at these tilt boundaries reveals a change in crystal orientation. One such region has been atomically zoomed into: Refer Fig. S1 (d).

In this work, we study the effect of tip-induced deformation on the strained moiré domains. The spatial extent of surface deformation induced by the tip depends on qualitative factors like tip apex geometry. To ensure the reproducible application of the same tip induced deformative forces on such various twist angled moiré, the entire scan area in Fig. S1 (a) is deformed in Fig. S1 (b).

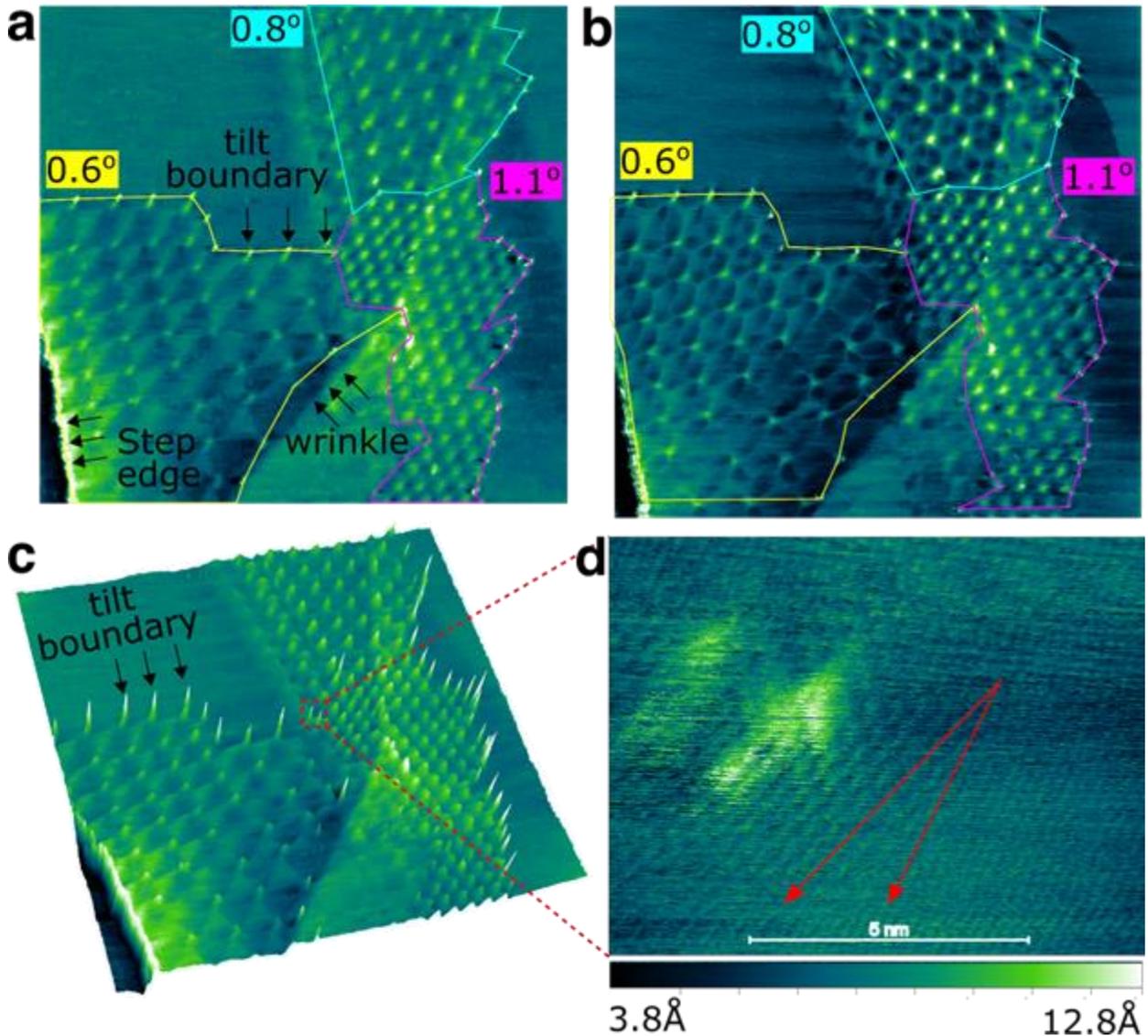

Figure **S1**: **Formation of multiple twist angle moiré pattern on graphite surface**
**(a)** Twist angle disorder is being exhibited where the presence of surface defects like wrinkles cause slight change in twist angle of the top carbon sheet. Moiré patterns are usually observed next to step heights. **(b)** The entire scan area in fig. S1 **(a)** is deformed at lower gap resistance which changes the moiré domain arrangement. **(c)** 3D view of fig. S1 **(a)** is shown to reveal that the moiré edges (or tilt boundaries) are topographically displaced higher than the moiré surface and are hence more strained. **(d)** A closer atomic image of the tilt boundary or edge is shown to reveal the change in crystallographic orientations around it as shown by red arrows.

**S2. Domain wall arrangement for various twist angles (θ) under deformation**

Irrespective of the twist angle and moiré material type, existence of certain domains like AA, hcp (ABA or AB or BA), fcc (ABC) and DWs remain consistent. With the twist angle reduction under the critical angle of 1°, there is amplification of the moiré wavelength that stretches the DWs and allows room for newer complex arrangement of the mentioned domains only. DWs are 1D structures sandwiched between the 2D domains. The strain energy balance between the two domains can be noticed from the change in shape, position and size of these DWs.

We observe the increment in moiré periodicity along the x axis in Fig. **S2 (a-e)** under the critical angle of 1° and observe a change in domain arrangement patterns from circular to triangular. AA domains slightly grow in size but move away from each other with θ reduction, stretching the DWs and allowing lateral growth of triangular ABC and ABA domains above the critical angle. A controversy (5,6) existed in relative assignment of ABC and ABA sites on moiré patterns due to possible electronic but studies (7) have shown that STM images of the moiré pattern are dominated by real deformations of the graphene layer, i.e., solely by the topography as also studied in Fig. 2 above. Following this theory, Fig. 3(e) indicates that ABA has a more stable stacking configuration than ABC (7–9) . The domains: AA, ABC, ABA and DW have also been identified (7) as ATOP, FCC, HCP and BRIDGE

On deforming all these angles as shown in the bottom row of Fig. **S2 (f-j)**, AAs appear pinned to their spatial location for all angles. Under critical angle, there is no change in domain arrangement. Above critical angle, ABA and ABC get compressed. All the moiré patterns in Fig. S2 have been deformed with the same tip to ensure the reproducible application of the same tip induced deformative forces that depend on qualitative factors like tip apex geometry. All the deformations on various twist angles have been done as a function of tunneling current at fixed sample bias of 100mV to eliminate any LDOS based electronic effects.

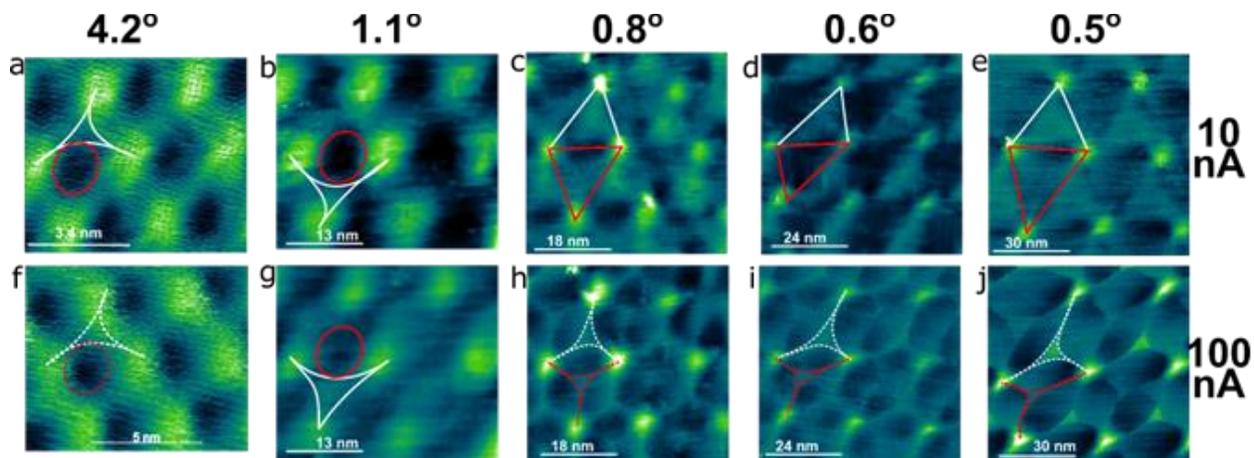

Figure **S2**: **Formation of multiple twist angle moiré pattern on graphite surface**
**(a-e)** Top row shows the effect of moiré domain arrangement with reduction in twist angle. The white lines are DWs enclosing the ABC (or rhombohedral) domains and the red lines enclose the stable ABA domains. The moiré wavelength along with ABC/ABA domain areas get enlarged along the x axis (θ reduction). Domain shape changes from circular (above 1°) to triangular (under 1°) accompanied with stretching along the length of DWs. **(f-j)** Bottom row shows the effect of domain arrangement under tip-induced deformation. There is no change in lateral arrangement under 1° but above 1° the ABC/ABA domains get compressed while there is stretch across the DWs. This strain energy balance shift is caused by the external tip induced strain.

## S3. Strain induced nematic phases on magic angle graphite moiré by STM topography

Deformation of moiré patterns under the critical angle have shown 6-fold symmetry which is distinct from moiré patterns above the critical angle of 1.1°. This makes studying the deformation of the magic angle moiré pattern interesting. We have observed by STM topography the existence of two kinds of deformation patterns on 1.1° graphite moiré as shown in Fig. S3. In the top row, we observe no change in deformation pattern and the three-fold symmetry is maintained. In the bottom row, we observe a striped pattern on deformation. The two moiré patterns in the top and bottom row are at different locations but of the same twist angle.

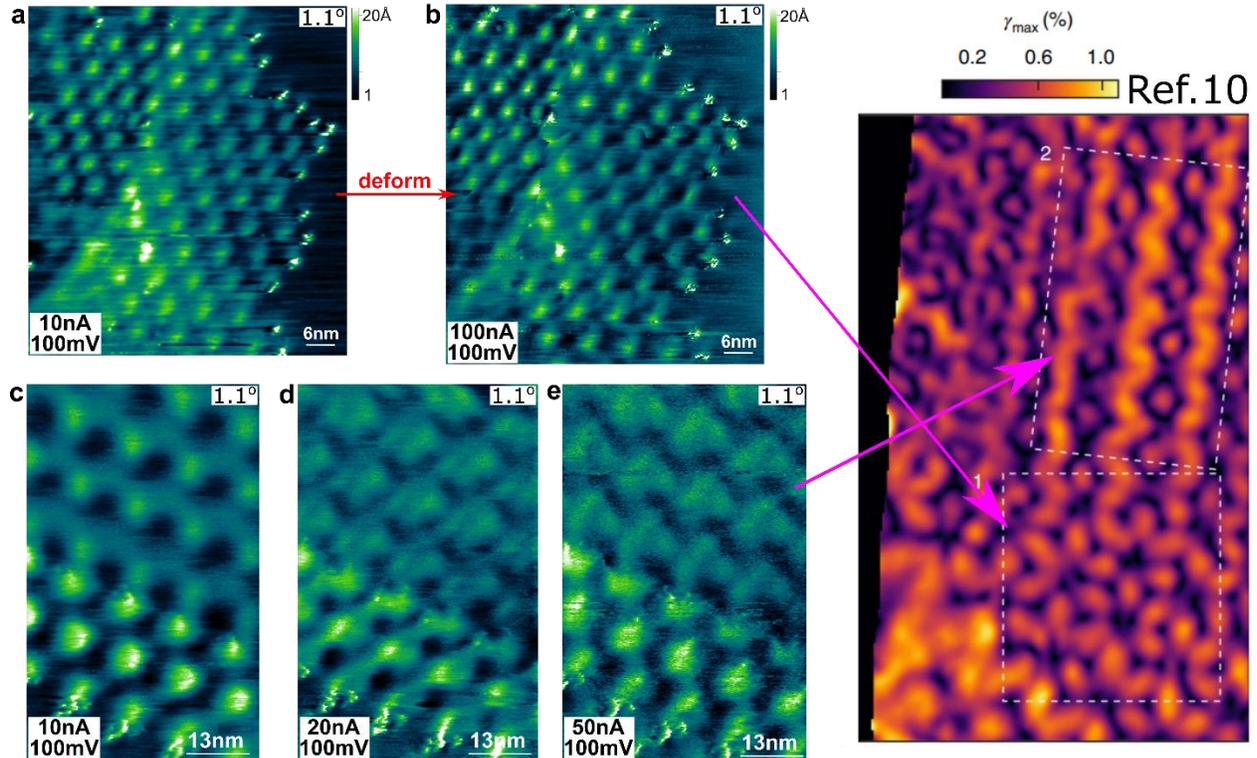

Figure **S3**: **Deformation of magic angle moiré exhibits two phases.**
**(a-b)** Top row shows the effect of tip-induced deformation on magic angle graphite moiré where the six fold symmetry remains intact even close to the tilt boundary and away from it. Another formed magic angle graphite moiré at a different location is shown in **(c-e)** where its deformation reveals the existence of a striped pattern near the tilt boundary and away from it. All the topographies have been measured at constant sample bias of 100mV.

Such dual pattern behavior has also been observed in magic angle TBG (10) where an applied hetero-strain accumulates anisotropically in DW regions generates distinctive striped strain phases. The uniaxial hetero-strain is not uniformly distributed on the edge of a moiré pattern. Strain field mapping has exhibited the 6-fold symmetric strain pattern being caused by 0.1% of hetero-strain whereas the striped pattern being caused by 1% of hetero-strain. However, we observed the existence of both the patterns near the edge of moiré where the hetero-strain is expected to be more.

Such mesoscale hetero-strain which causes the anisotropic amplification and deformation of DW into striped phase (in Fig. S3(e) ) has localized, symmetry-breaking nanoscale features and it may be linked to nematic phases also observed STM in magic angle TBG (11) . The symmetry breaking from 6 fold symmetry to striped phase has been observed via LDoS mapping depending on electron doping with the strongest nematicity near half-filling. Such strain induced nematic susceptibility has also been observed in iron pnictides (12) . The interaction of such ordered states with superconductivity in TBG remains to be investigated.